\documentstyle[12pt]{article}
\begin{document}
\title{
Stability of  rotating supermassive stars in presence of dark matter
background}

\author{G.S. Bisnovatyi-Kogan,
\thanks{ Space Research Institute,
Russian Acadamey of Sciences, Moscow, Russia. Email: gkogan@mx.iki.rssi.ru}}
\date{}
\maketitle
\begin{abstract}
Stability of supermassive stars embedded in hot dark matter is
investigated on the base of the energetic method. Stability effect of dark
mater is compared with rotational stabilization and preference of the
last factor is advocated.
\end{abstract}

\section{Introduction}
Supermassive stars, considered hystoricaly as a first model of quasars
and active galactic nuclei (AGN) suffer from early instability to
relativistic collapse, imposing their short life time (Fowler, 1964;
Zeldovich and Novikov, 1965; Ozernoy, 1966).
This instability is connected with a high
entropy of the matter in such stars, what leads to prevailing of radiation
pressure, closeness of the adiabatic index to 4/3 and small
stability resource.

A common way to overcome this instability is to consider rotating
superstars, what may postpone the moment of collapse to $3 \times 10^4$
years for solid body rotation with angular momentum and mass losses
(Bisnovatyi-Kogan, Zeldovich and Novikov, 1967), and much longer for a
differentially rotating star evolving with almost constant angular
momentum (Fowler, 1966; Bisnovatyi-Kogan and Ruzmaikin, 1973).
Farther developement of a theory and collection of observations led to
accreting supermassive black hole (Lynden-Bell, 1969) as a most promising
model for quasars and AGN. Nevertheless, formation of supermassive stars
on early stages of the Universe expansion, their loss of stability
with subsequent collapse or explosion (Bisnovatyi-Kogan, 1968; Fricke, 1973;
Fuller et al, 1986) could be important for early formation
of heavy elements, observed in spectra of the most distant objects with
red shift $\sim 5$, creation of perturbations for large scale structure
formation, influence on small scale fluctuations of microwave background
radiation (Peebles, 1987; Cen et al, 1993). A necessity of a presence of a
dark matter in modern cosmological models makes it important to include
it into stability analysis of supermassive stars. This was done by
McLaughlin and Fuller (1996), who dealed with nonrotating superstars.
Here we consider the same problem for rotating superstars,
using energetic method,
which is asymptotically exact for supermassive stars (Fowler, 1964;
Zeldovich and Novikov, 1967). We show that
rotational effects occure to be more important
for realistic choice of parameters.

\section{Stability analysis}

In supermassive stars with equation of state $P=P_r+P_g=\frac{aT^4}{3}+
\rho {\cal R}T$ there is $P_r \gg P_g$ due to high entropy of such stars.
Besides, such stars are fully convective and
entropy is uniform over them, so the spatial structure is
well described by a polytropic distribution, corresponding to $\gamma=4/3$.
In the energetic method, considered by Fowler (1964), Zeldovich and Novikov
(1965), and formalized by Bisnovatyi-Kogan (1966), the spatial profile of
the star is fixed during perturbations $\rho = \rho_c \phi(m/M)$, and
stability analysis is reduced to the set of algebraic equations. When the
main terms in total energy (without a rest mass) are the energy of
radiation and newtonian gravity, small terms connecting with plasma energy,
general relativistic corrections, rotation and a dark matter, are calculated
using polytropic, with $\gamma=4/3$, density distribution, corresponding to
isentropic star with pure radiation pressure. In this situation influence of
a hot dark matter, which density does not change during perturbations,
should be taken into account by account of a newtonian gravitational
energy of the star in the dark matter potential, because GR effects of a
dark matter are of a higher order of magnitude (McLaughlin, Fuller, 1996).

For radiation dominated plasma there is a following expression for the
adiabatic index, determining the stability to a collapse

\begin{equation}
\label{ref1}
 \gamma=\left(\frac{\partial\log P}{\partial \log\rho}\right)_S \approx
 \frac{4}{3}
 \left(1+\frac{\cal R}{2S}\right)=\frac{4}{3}+\frac{\beta}{6},
 \end{equation}
 where $\beta=\frac{P_g}{P}= \frac{4{\cal R}}{S}$.
 In the radiationaly dominated supermassive
 star there is a unique connection between its mass $M$ and entropy per
 unit mass $S$ (Zeldovich and Novikov, 1967)

\begin{equation}
\label{ref2}
 M=4.44 \left(\frac{a}{3G}\right)^{3/2}\left(\frac{3S}{4a}\right)^2,
 \end{equation}
where $a$ is a constant of the radiation energy density, and numerical
coefficient is related to the polytropic density distribution with $\gamma=
4/3$. At the point of a loss of stability the critical value of an
average adiabatic index
$<\gamma>$ in selfgravitating nonrotating star with account of post-newtonian
corrections is determined by a relation (Zeldovich and Novikov, 1967)

\begin{equation}
\label{ref3}
<\gamma>_{crs}=\frac{4}{3}+ \delta_{GR}=
\frac{4}{3}+ \frac{2}{3} \frac{\varepsilon_{GR}}
{\varepsilon_N} \approx \frac{4}{3}+ 0.99 \frac{GM^{2/3} \rho_c^{1/3}}{c^2}.
\end{equation}
Here averaging is done over the stellar volume with a pressure $P$ as
a weight function, $\varepsilon_{GR}$ and  ${\varepsilon_N}$ are
post-newtonian and newtonian gravitational energies of the superstar,
respectively. From comparison between (\ref{ref1}) and (\ref{ref3})
we get a well known relation for a critical central density of a
supermassive star stabilized by plasma

\begin{equation}
\label{ref4}
\rho_c=0.10\frac{{\cal R}^3 c^6}{G^{21/4} a^{3/4}} M^{-7/2} \approx
1.8 \times 10^{18} \left(\frac{M_{\odot}}{M}\right)^{7/2} {\rm g/cm}^3.
\end{equation}
Here and below we consider for simplicity a pure hydrogen plasma.
Newtonian energy of a superstar $\varepsilon_{nd}$ in the gravitational field of uniformly
distributed dark matter with a density $\rho_d$ is written as

\begin{equation}
\label{ref5}
\varepsilon_{nd}=\int_0^M \varphi_d dm.
\end{equation}
The gravitational potential of a uniform dark
matter $\varphi_d$ is written as

\begin{equation}
\label{ref6}
\varphi_d=\frac{2 \pi}{3}G\rho_d r^2-\frac{3}{2}\frac{GM_d}{R_d},
\end{equation}
where $R_d$ is much larger then stellar radius $R$, and $M_d$ is a total mass
of the dark matter halo.
Stability does not depend on normalization of the gravitational potential
so we shall omit the constant value in (\ref{ref5}).
It follows from (\ref{ref5}) and (\ref{ref6}) that during variations
$\varepsilon_{nd} \sim \rho_c^{-2/3}$, while
$\varepsilon_{GR} \sim \rho_c^{2/3}$  and
$\varepsilon_{N} \sim \rho_c^{1/3}$ (Zeldovich and Novikov, 1967).
Critical value of the adiabatic index in the energetic method is
obtained in the most convenient way by equating to zero a second derivative
of the total energy over $\rho^{1/3}$. As a result for nonrotaing superstar
in presence of a dark matter the critical value of an average adiabatic
index $<\gamma>_{crnrot}$ is determined by

\begin{equation}
\label{ref7}
<\gamma>_{crnrot}= <\gamma>_{crs} + \delta_{dm}=
\frac{4}{3}+ \frac{2}{3} \frac{\varepsilon_{GR}}
{\varepsilon_N} - 2 \frac{\varepsilon_{nd}}{\vert\varepsilon_N\vert}.
\end{equation}
Taking into account, that $\varepsilon_N \approx 12 \pi\int_0^R P r^2 dr$
and (\ref{ref5}) - (\ref{ref7}) we obtain the value of a dark matter
correction to the stability criteria $\delta_{dm}$
coinsiding with
the corresponding correction $\delta_{dm}$ in the formula (11e)
from McLaughlin and Fuller (1996) (later with subscript $_{MF}$:
(11e$_{MF})$). The relation for a critical density
in presence of a dark matter is obtained by comparison of
(\ref{ref1}) and (\ref{ref7}). We get

 \begin{equation}
 \label{ref8}
 0.99\frac{GM^{2/3} \rho_c^{1/3}}{c^2}=
 4.1\frac{\rho_d}{\rho_c}+\frac{\beta}{6}=
 4.1\frac{\rho_d}{\rho_c}+
 \frac{\cal R}{2a}\left(\frac{4.44}{M}\right)^{1/2}
 \left(\frac{a}{3G}\right)^{3/4}.
 \end{equation}
 Here relations $R=\frac{M^{1/3}\rho_c^{-1/3}}{0.426}$ and
 $\int_0^R \rho r^4 dr=6.95 \times 10^{-4} \rho_c R^5$, based on Emden
 polytropic distribution with $\gamma=4/3$, were used (see Bisnovatyi-Kogan,
 1989). Relation (\ref{ref8}) is reduced to
 $(6_{MF})$, note a misprinting in the sign in $(6_{MF})$.

\begin{equation}
\label{ref9}
2.8 \times 10^{-3}\left(\frac{M}{M_6}\right)^{2/3} \rho_c^{4/3}=
3.5 \times 10^{-4}\left(\frac{M_6}{M}\right)^{1/2} \rho_c+\rho_d.
\end{equation}
In Fig.1 solution of (\ref{ref9}) is presented, as in Fig.2$_{MF}$.

\section{Stability of rotating stars}
Rotaion was shown to be a main stabilization factor of a superstar. Let us
consider a rigid rotation, when its energy is a small correction to
the energy of radiation and the energetic method is a good approach.
When losses of an angular momentum during evolution are negligible we
distinguish between rapidly rotating (RR) and slowly rotating (SR)
superstars. In RR case a superstar reaches the state of rotational
equatorial breaking before loosing its dynamical instability, and
in SR case instability comes first. If a superstar has an angular momentum
$J$, then its rotational energy $\varepsilon_{rot} \approx 1.25
J^2\rho_c^{2/3}M^{-5/3}$, and a ratio $\varepsilon_{rot}/\varepsilon_{GR}$
remains constant during evolution
(Bisnovatyi-Kogan, Zeldovich,
Novikov, 1967).
In presence of rotation and dark matter the critical
value of the adiabatic index $<\gamma>_{crrot}$ is written as

\begin{equation}
\label{ref10}
<\gamma>_{crrot}=
\frac{4}{3}+ \frac{2}{3} \frac{\vert\varepsilon_{GR}\vert-\varepsilon_{rot} }
{\vert\varepsilon_N\vert} -
2 \frac{\varepsilon_{nd}}{\vert\varepsilon_N\vert},
\end{equation}
and the relations for determination of a critical central density, instead
of (\ref{ref9}), is written as

\begin{equation}
\label{ref11}
2.8 \times 10^{-3}\left(\frac{M}{M_6}\right)^{2/3} \rho_c^{4/3}
\left(1-\frac{\varepsilon_{rot}}{\vert\varepsilon_{GR}\vert}\right)=
3.5 \times 10^{-4}\left(\frac{M_6}{M}\right)^{1/2} \rho_c+\rho_d.
\end{equation}
As follows from (\ref{ref11}), a superstar does dot loose its stability
when $\varepsilon_{rot} > \vert\varepsilon_{GR}\vert $. This qualitative
result, obtained in the post-newtonian approximation, remains to be valid
in a strong gravitational field and reflects
a presence of a limiting specific angular momentum $a_{lim}=GM/c$,
so that a black holes with a Kerr metric may exist only at $a< a_{lim}$
(Misner, Thorne, Wheeler, 1973).

A RR superstar in a course of the evolution reaches instead a limit of
a rotational instability, and equatorial mass shedding begins, leading to
a loss of an angular momentum. Such star will loose the stability when the
anuglar momentum will become less then the limiting value. The stage of a
mass loss was examined by
(Bisnovatyi-Kogan, Zeldovich, Novikov, 1967), where it was shown that
this stage may last about 10 times longer, then a maximum evolution time
to approach the rotational instability point.
RR star reaches the stage of a rotational instability at different central
densities, depending on $J$, but the ratio of rotatonal and Newtonan
gravitational energy on the mass-shedding curve remains constant
(Bisnovatyi-Kogan, Zeldovich, Novikov, 1967)
\footnote{Note, that
in presence of a dark matter halo mass outflow begins at rotational
energy approximately $(1+54\rho_d/\rho_c)$ times larger,
then in (\ref{ref12}) due to additional
gravity of a dark matter. This correction is small for a considered halo
density on the mass-shedding curve at $J>J_0$ (see below), and is
neglected in this section.}

\begin{equation}
\label{ref12}
\varepsilon_{rot}=0.00725\vert\varepsilon_{N}\vert.
\end{equation}
The energy of a rotating supertar in equilibrium in presence of a
hot dark matter may be written as

\begin{equation}
\label{ref13}
\varepsilon_{eq}=
-\varepsilon_{gas}+\vert\varepsilon_{GR}\vert
-\varepsilon_{rot}+3\varepsilon_{nd}.
\end{equation}
In the main term for a superstar in equilibrium  a relation is valid

\begin{equation}
\label{ref14}
\varepsilon_{gas}=\frac{\beta}{2}\vert\varepsilon_{N}\vert.
\end{equation}
Taking into account (\ref{ref12}), (\ref{ref14}), we get an expression for
an equilibrium energy along the mass-shedding curve (with variable $J$)

\begin{equation}
\label{ref15}
\varepsilon_{eq}=
-\left(0.00725+\frac{\beta}{2}\right)\vert\varepsilon_N\vert+
\vert\varepsilon_{GR}\vert+3\varepsilon_{nd}.
\end{equation}
The curve $\varepsilon_{eq}(\rho_c)$ has a minimum at the central density,
determined by a relation

\begin{equation}
\label{ref16}
2.8 \times 10^{-3}\left(\frac{M}{M_6}\right)^{2/3} \rho_c^{4/3} =
3.5 \times 10^{-4}\left(\frac{M_6}{M}\right)^{1/2} \rho_c+
5.9 \times 10^{-4}\rho_c+\rho_d.
\end{equation}
From comparison (\ref{ref16}) and (\ref{ref11}) with account of (\ref{ref12})
it is clear, that
dynamical instability cannot occure in the minimum of the mass-shedding
curve, and after crossing it the evolution proceeds with a substantial
mass and angular momentum losses. Central density of the superstar
in the minimum of the mass-
shedding curve (\ref{ref15})
with and without dark matter are represented in the fig.1.

Let us find a parameters of a superstar, at which its critical state
is situated on the mass-shedding curve. These parameters satisfy
sumultanously the relations (\ref{ref11}) and (\ref{ref12}),
what leads to the equation for determination of a central density
in the form

\begin{equation}
\label{ref17}
2.8 \times 10^{-3}\left(\frac{M}{M_6}\right)^{2/3} \rho_c^{4/3} =
3.5 \times 10^{-4}\left(\frac{M_6}{M}\right)^{1/2} \rho_c+
12 \times 10^{-4}\rho_c+\rho_d.
\end{equation}
The relation (\ref{ref12}) is used for determination of
an angular momentum of the superstar $J=J_0$ with $\rho_c$ from
(\ref{ref16}) and $J=J_1<J_0$ with $\rho_c$ from (\ref{ref17}).
Solution of this equation is also given in fig.1. As may be seen from fig.1,
the stabilizing effect of rotation on the mass-shedding curve
at $J=J_1$ is more
important, then stabilization by a hot dark matter, which influence
decreases at increasing of a central density. With
account of a longer state
of the evolution with mass loss until reaching the point of the loss
of stability at larger $J$,
this conclusion becomes even stronger.

Determination of the central density at the point of the loss of stability
is often not enough for characterizing the stabilization. The
effect of stabilization is characterized by increase of a life time $\tau$
of the star until it reaches the point of a loss of stability. In turn
$\tau$ depends on the energy loss rate and changing of a binding energy
of a superstar along its evolutionary path.

\section{Evolution of a superstar}

In massive superstars the temperature and the density at the point of
a loss of stability are not enough for sufficient nuclear energy production,
so gravitational energy is the main source. Luminosity of a superstar
is determined by its mass and is very close to the Eddington critical
luminosity
$L_E=\frac{4\pi cGM}{\kappa}$
(Fowler, 1964; Zeldovich and Novikov, 1965)
In rotating superstars mass loss begins at
luminosity less then $L_E$ due
to the action of a centrifugal force, and the presence of a dark matter
imply increasing of the luminosity of a superstar, because the radiation
force must balance the gravity of the dark matter in addition to the
selfgravity of a superstar. When using $L_E$ for estimation of
the evolution time we should take into account possible mass loss
in the case of a rigid rotation.
In the presence of a dark matter we would overestimate the lifetme of
the superstar by using (\ref{ref18}).

Taking into account the gravity of a dark matter, we obtain that
radiational force balance the total gravity at $L=L_{dE}$

\begin{equation}
\label{ref19}
L_{dE}=\frac{4\pi cGM}{\kappa}\left(1+54.2\frac{\rho_d}{\rho_c}\right).
\end{equation}

Nonmonotonial dependence of the binding energy
$\varepsilon_b=-\varepsilon_{eq}$ on the central density may confuse
the estimation of a stabilization effect, using only a central density
data. This is what takes place along the mass-shedding curve. Here
maximum of the binding energy corresponds to the central density, determined
by (\ref{ref16}), and at larger central density on this curve
in the point coinciding with the critical point,
determined by (\ref{ref17}),
the binding energy is less, so the evolution time is smaller.
Comparison of binding energies in these two points is easy do to
analytically for small dark matter influence.
For minimum of mass-shedding curve we get a central density from
(\ref{ref16}) as

\begin{equation}
\label{ref20}
\rho_{cms}^{1/3}=
A\frac{\rho_c^{1/3}\varepsilon_N}{\varepsilon_{GR}}+
\frac{3}{A^3}\frac{\rho_c^{1/3}\varepsilon_{nd}\varepsilon_{GR}^2}
{\varepsilon_N^3}=
\end{equation}
$$0.21\left(\frac{M_6}{M}\right)^{2/3}+
0.12\left(\frac{M_6}{M}\right)^{7/6}+  $$
$$3.7\times 10^4 \rho_d \left(\frac{M}{M_6}\right)^{4/3}
\left[1+0.58 \left(\frac{M_6}{M}\right)^{1/2}\right]^{-3}.$$
At the critical point on this curve it follows from (\ref{ref17})

\begin{equation}
\label{ref21}
\rho_{ccs}^{1/3}=
(2A-\frac{\beta}{4}) \frac{\rho_c^{1/3}\varepsilon_N}{\varepsilon_{GR}}+
\frac{3}{(2A-\frac{\beta}{4})^3}
\frac{\rho_c^{1/3}\varepsilon_{nd}
\varepsilon_{GR}^2}{\varepsilon_N^3}=
\end{equation}
$$0.42\left(\frac{M_6}{M}\right)^{2/3}+
0.12\left(\frac{M_6}{M}\right)^{7/6}+$$
$$4.7\times 10^3 \rho_d \left(\frac{M}{M_6}\right)^{4/3}
\left[1+0.29 \left(\frac{M_6}{M}\right)^{1/2}\right]^{-3},$$
where  $A=\frac{1}{2}(0.0075+\frac{\beta}{2})$.
The binding energy in the first point, with account of (\ref{ref20}) is

\begin{equation}
\label{ref22}
\varepsilon_{bms}=
\frac{1}{2}\left(0.00725+\frac{\beta}{2}\right)\vert\varepsilon_N\vert
-6\varepsilon_{nd}=
\end{equation}
$$A^2\frac{\varepsilon_N^2}{\varepsilon_{GR}}
-\frac{3}{A^2}\frac{\varepsilon_{nd}
\varepsilon_{GR}^2}{\varepsilon_N^2},$$
and in the second point, with account of (\ref{ref21}) is

\begin{equation}
\label{ref23}
\varepsilon_{bcs}=
\frac{\beta}{4}\vert\varepsilon_N\vert
-6\varepsilon_{nd}=
\end{equation}
$$(2A-\frac{\beta}{4})\frac{\beta}{4}
\frac{\varepsilon_N^2}{\varepsilon_{GR}}-
\frac{3(4A-\frac{3}{4}\beta)}{(2A-\frac{\beta}{4})^3}
\frac{\varepsilon_{nd}\varepsilon_{GR}^2}{\varepsilon_N^2}.$$
We get from comparison of (\ref{ref22}) and (\ref{ref23})

\begin{equation}
\label{ref24}
\varepsilon_{bms}-\varepsilon_{bcs}=
(A-\frac{\beta}{4})^2 \frac{\varepsilon_N^2}{\varepsilon_{GR}}-
6\frac{(2A-\frac{\beta}{8})(A-\frac{\beta}{4})^2}{A^2(2A-\frac{\beta}{4})^3}
\frac{\varepsilon_{nd}\varepsilon_{GR}^2}{\varepsilon_N^2}.
\end{equation}
The second term in (\ref{ref24}) is much smaller then the first, so the
minimum point with smaller central density (\ref{ref19}) has larger binding
energy then the point of loss of stability (\ref{ref20}).
Time of the evolution of a superstar until the point of the loss
of stability $\tau$ is equal to

\begin{equation}
\label{ref25}
\tau=\int_0^{\rho_{cc}}\frac {d\varepsilon_b}{L_E}
\end{equation}
Due to increase of the binding energy with increase of a central density
the evolution becomes slower with time. In the minimum of the mass-
shedding curve the central density for $M=10^6 M_{\odot}$ is about
$0.06$ g/cm$^3$ (see Fig.1), so the increase of the critical luminosity due
to the gravity of a dark matter is negligible for $\rho_d=10^{-5}$ g/cm$^3$,
see (\ref{ref19}). It means, that estimations of the life times of the
superstar made by Bisnovatyi-Kogan, Zeldovich, Novikov (1967) for
rotating superstars without a presence of a dark matter remain valid
for parameters of the dark matter,
considered above (McLaughlin, Fuller, 1996).
The maximal evolution time of a rigidly rotating superstar without a
mass loss cannot exceed $3\times 10^3$ years, and the stage of a mass
loss until reaching the point of a loss of stability could be an
order of a magnitude longer.

\bigskip
{\bf Acknowledgements} \\
This work was partly supported by
NSF grant AST-9320068, Russian
Fundametal Research Foundation
grant No. 96-02-16553 and grant of a Ministry of Science and Technology
1.2.6.5.

\bigskip
\begin{figure}
\caption
{The correction terms  $\delta_{GR}$ and $\delta_{dm} + \delta_{GR}$,
the quantities $\beta$/6 (line a),
$\beta$/6+$(\varepsilon_{rot}/\varepsilon_N)_{sh}$/3 (line b), and
2$\beta$/6+
$(\varepsilon_{rot}/\varepsilon_N)_{sh}$/3 (line c),
 as functions of the central density of a
supermassive star with $M=10^6 M_{\odot}$, and dark matter density
of $10^{-5}$ g/cm$^3$. The instability points for nonrotating star
occure at intersection of the
correction term curves with the line a. Mass shedding in the stable
star with angular momentum $J_0$ (see text) occures at intersection of
correction term curves with the line b, and critical point on the
mass-shedding curve is determined by a corresponding intersection with
the line c.}
\label{fig1}
\end{figure}
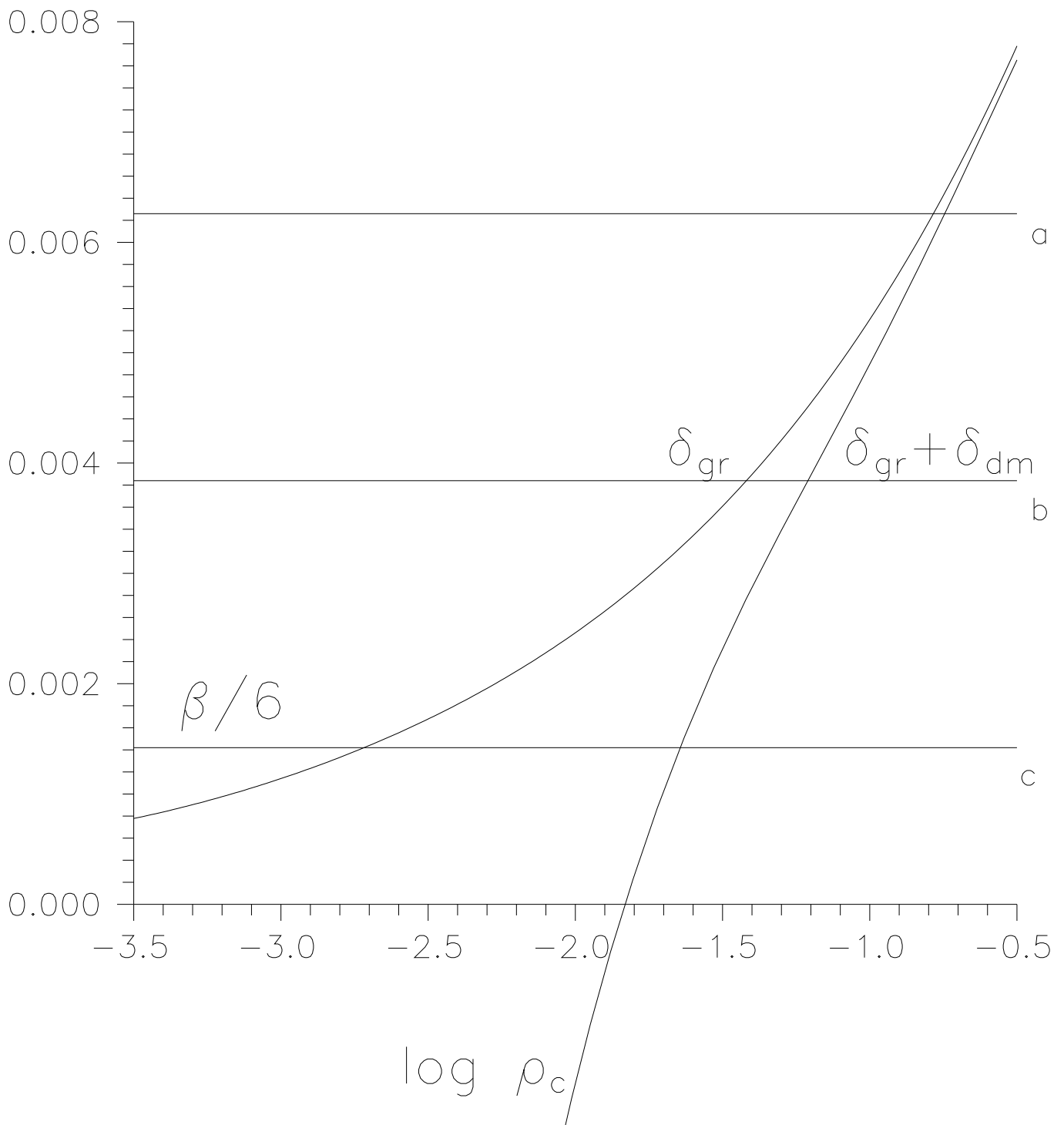
\end{document}